\documentclass[english]{article}
\usepackage[latin9]{inputenc}
\usepackage{amsmath}
\usepackage{graphicx}
\providecommand{\tabularnewline}{\\}
\usepackage{babel}
\usepackage{subfig}
\usepackage{caption}
\usepackage{stackengine}
\usepackage{float}
\makeatletter
\begin{document}

\title{CASCADE CALCULATION WITH SCHEMATIC INTERACTIONS}

\author{Levering Wolfe and Larry Zamick \\
 Department of Physics and Astronomy, \\
 Rutgers University, Piscataway, New Jersey 08854}

\maketitle

\section{Abstract}

In previous works we considered schematic Hamiltonians represented
by simplified matrices. We defined 2 transition operators and calculated
transition strengths from the ground state to all exited states.In
many cases the strengths decreased nearly exponentially with excitation
energy. Now we do the reverse We start with the highest energy state
and calculate the cascade of transitions until the ground states is
reached.On a log plot we show the average transition strength as a
function of the number of energy intervals that were crossed. We give
an analytic proof of exponential behavior for transition strength
in the weak coupling limit for the T$_2$ transition operator.

\section{Introduction}

In ref {[}1{]} and {[}2{{]} }A.Kingan et al. calculated M1 transition
from the lowest J=1+ state of a selected even-even nucleus to excited
states. In the first work they emphasized strong transitions to individual
states.. Although transitions to the ground state were large (inverted
scissors) there were many other transition strengths that were even
larger i.e to other J=0+ states and to many J=2+states. Basically
they were calculating double magnetic dipole strength.

In Ref {[}2{]} transitions to nearly all states we're considered-strong
or weak. A wide spread of results was noted. However when the transition
strengths in certain energy intervals (bins)were summed up one saw
an exponential decrease in strength with excitation energy. Coincidentally
a similar behavior, exponential decrease with excitation energy, was
also found in a different problem - a study of schematic matrix hamiltonians
for their own sake{[}3,4,5{]}.The study was mainly on tridiagonal
matrices{[}3,4{]} but also later pentadiagonal were considered{[}5{]}.
We will here consider also heptadiagonal matrices.

We were informed about experiments and calculations which do the opposite
of what we have done-start with the highest level and calculate all
transition strengths as the nucleus deexcites emitting gamma rays
{[}6-11{]}. They plot the average M1 strength versus gamma ray energy. They find a strong increase in average
M1 strength as one goes to low gamma ray energies.

We will here also do such calculations with our schematic Hamiltonian
matrices. Because they are simpler perhaps we can cast some light
on what is causing the behaviors that have been observed. Indeed in
ref {[}5{]} we were able to get some analytic results for wave functions
and transitions from the ground state, especially in the weak coupling
limit. in contrast to [6] and [8-12] we present our results as log plots. Brown and Larsen also used a
log plot [7]. It should be made clear that our basic aim is to study the properties
of the matrix Hamiltonian of Table 1. We will not attempt to fit the
results of {[}6-12{]} but we will see if there are some common features
that might be of interest.

In a previous work [5] with tridiagonal matrices it was shown that in the weak coupling limit
the ground state column vector (a$_0$,a$_1$...a$_n$....) was such that a$_n$=$(\frac{-v}{E})^n \frac{1}{n!} $. We recognized
this as an a expansions in a Taylor series of e$^{-\frac{v}{E}} $. The $\frac{1}{n!} $ terms come from energy
denominators in the wave function

$\Psi$= $\varPhi$ + $\frac{1}{E_0-H_0} $QV$\varPsi$ = (1+$\frac{1}{E_0-H_0} $QV +$\frac{1}{E_0-H_0} $QV$\frac{1}{E_0-H_0} $QV+..... )$\varPhi$

\section{The Calculation.}

We here show a 11 by 11 Hepta diagonal matrix we will be working with.
We can also get results for tri and penta diagonal matrices by setting
certain parameters to zero.

\begin{table}[H]
\captionof{table}{The septa- diagonal matrix} %
\begin{tabular}{|c|c|c|c|c|c|c|c|c|c|c|}
\hline 
0  & v  & w  & x  &  &  &  &  &  &  & \tabularnewline
\hline 
v  & E  & v  & w  & x  &  &  &  &  &  & \tabularnewline
\hline 
w  & v  & 2E$ $  & v  & w  & x  &  &  &  &  & \tabularnewline
\hline 
x  & w  & v  & 3E  & v  & w  & x  &  &  &  & \tabularnewline
\hline 
 & x  & w  & v  & 4E  & v  & w  & x  &  &  & \tabularnewline
\hline 
 &  & x  & w  & v  & 5E  & v  & w  & x  &  & \tabularnewline
\hline 
 &  &  & x  & w  & v  & 6E  & v  & w  & x  & \tabularnewline
\hline 
 &  &  &  & x  & w  & v  & 7E  & v  & w  & x\tabularnewline
\hline 
 &  &  &  &  & x  & w  & v  & 8E  & v  & w\tabularnewline
\hline 
 &  &  &  &  &  & x  & w  & v  & 9E  & v\tabularnewline
\hline 
 &  &  &  &  &  &  & x  & w  & v  & 10E\tabularnewline
\hline 
\end{tabular}
\end{table}

We will be considering and comparing results for tridiagonal, pentadiagonal
and heptadiagonal matrix Hamiltoninans. For tridiagonal case, we set
x=0, w=0 and for the pentadiagonal case, x=0 in the matrix shown in
Table 1. In previous works {[}3,4,5{]} we defined 2 types of transition
operators \textless{}n T$_1$(n+1)\textgreater{}= 1 and \textless{} n
T$_2$ (n+1) \textgreater{} = $\sqrt{n+1}$. In this work we will only
show results for the latter.

\section{The Figures}

We here present several figures. We only include the T$_2$ coupling case.
We show side by side figures of the cascade down average strength
(left) and the transition strength up (right). This is done first
for the tridiagonal case with v=0.01, 0.1 and 1. Next for the petadiagonal
case w=v(0.01, 0.1). Next heptadiagonal with v=w=x(0.01, 0.1, 1) and
then again heptadiagonal w=$\frac{v}{2\!}$,x=$\frac{v}{3\!}$ (v=0.01,
0.1 and 1). Note that although not identical the cascade (down) figures
look very similar to the corresponding excitation (up) figures.

\begin{center}
    
\begin{figure}[!htbp]
\footnotesize
\stackunder[5pt]{\includegraphics[width=6cm,height=3.66cm]{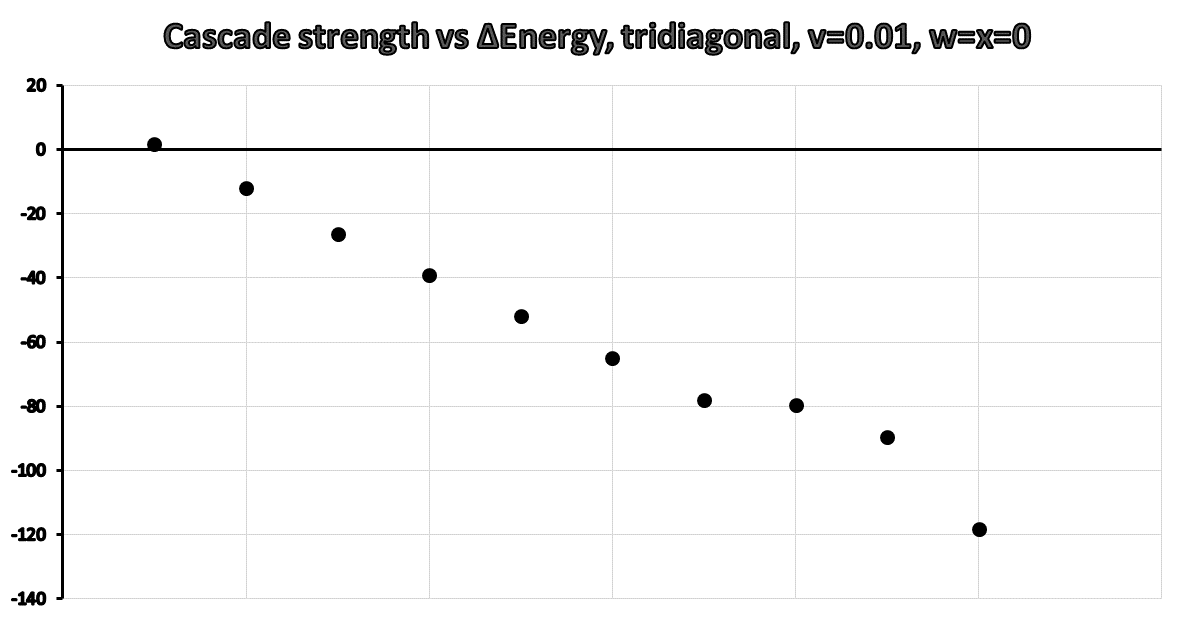}}{ }%
\hspace{1cm}%
\stackunder[5pt]{\includegraphics[width=6cm,height=3.66cm]{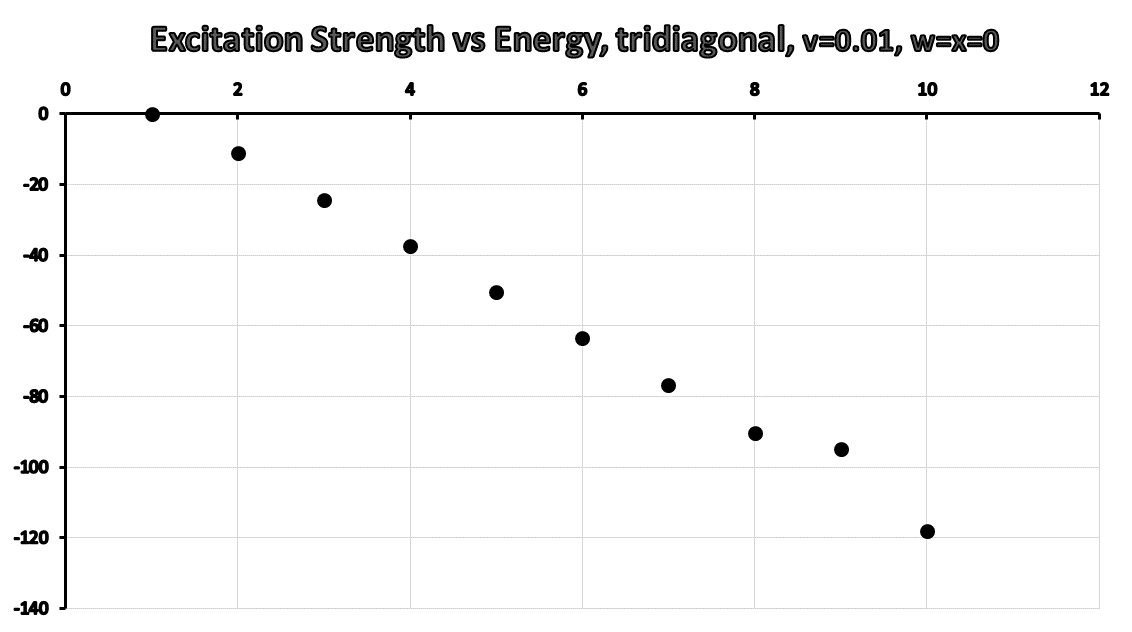}}{}
\caption*{}
\end{figure}

\end{center}

\begin{center}
    
\begin{figure}[!htbp]
\footnotesize
\stackunder[5pt]{\includegraphics[width=6cm,height=3.66cm]{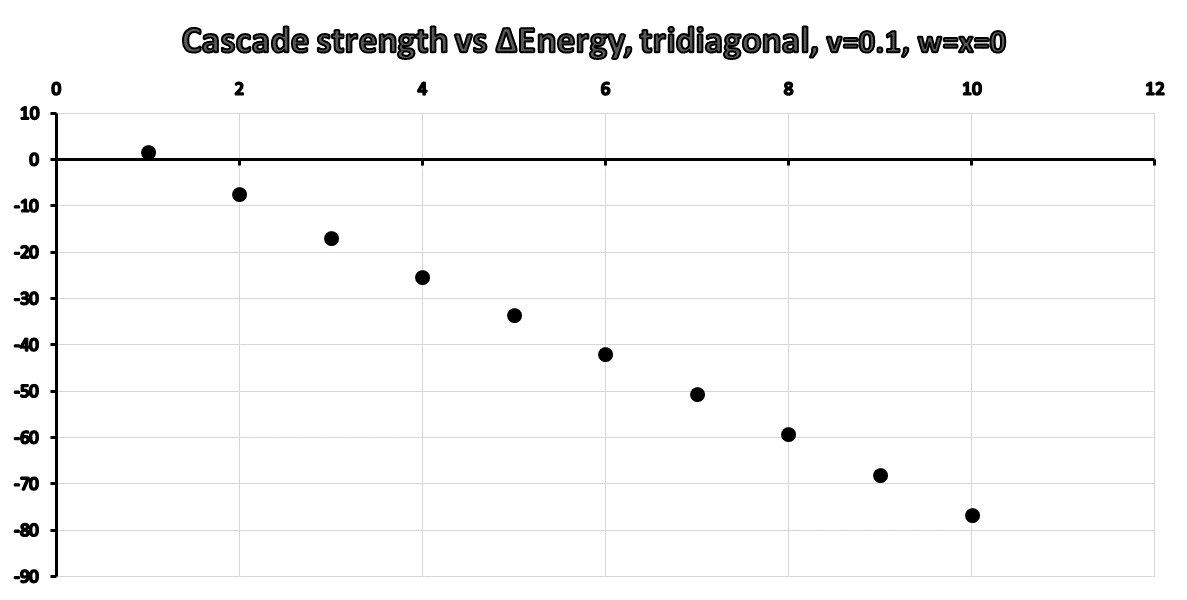}}{ }%
\hspace{1cm}%
\stackunder[5pt]{\includegraphics[width=6cm,height=3.66cm]{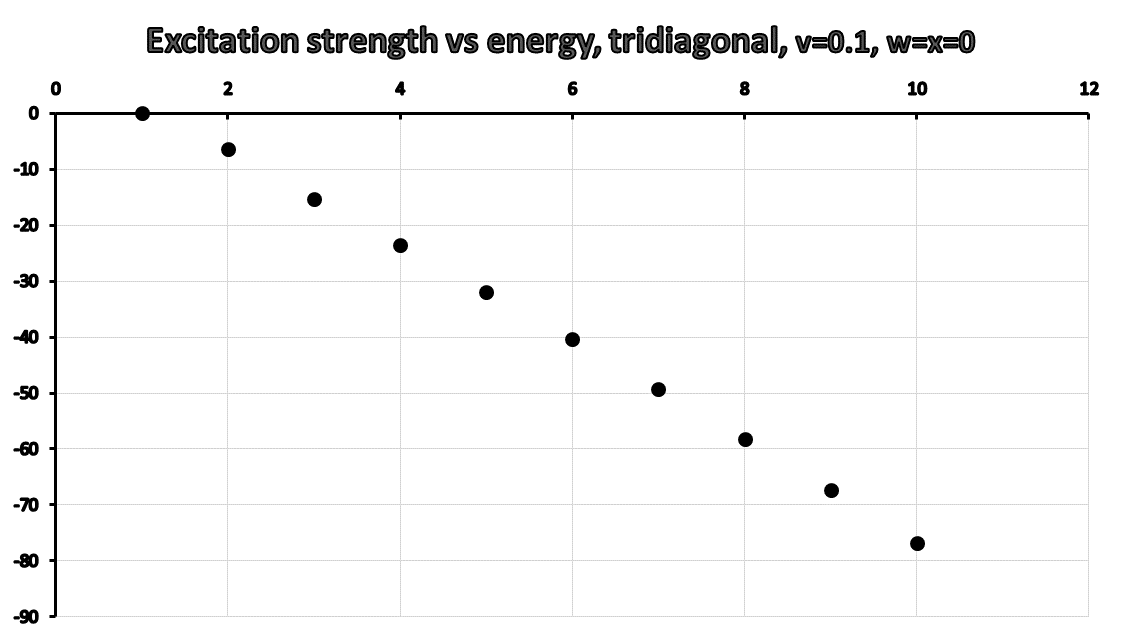}}{}
\caption*{}
\end{figure}

\end{center}

\begin{center}
    
\begin{figure}[!htbp]
\footnotesize
\stackunder[5pt]{\includegraphics[width=6cm,height=3.66cm]{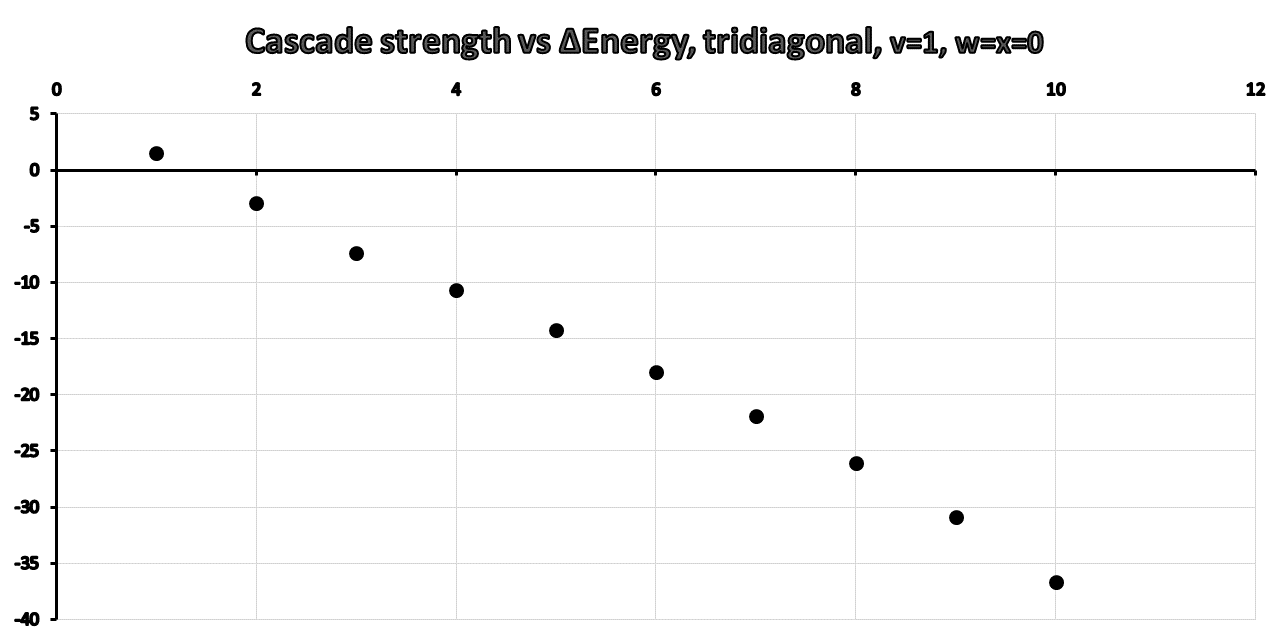}}{ }%
\hspace{1cm}%
\stackunder[5pt]{\includegraphics[width=6cm,height=3.66cm]{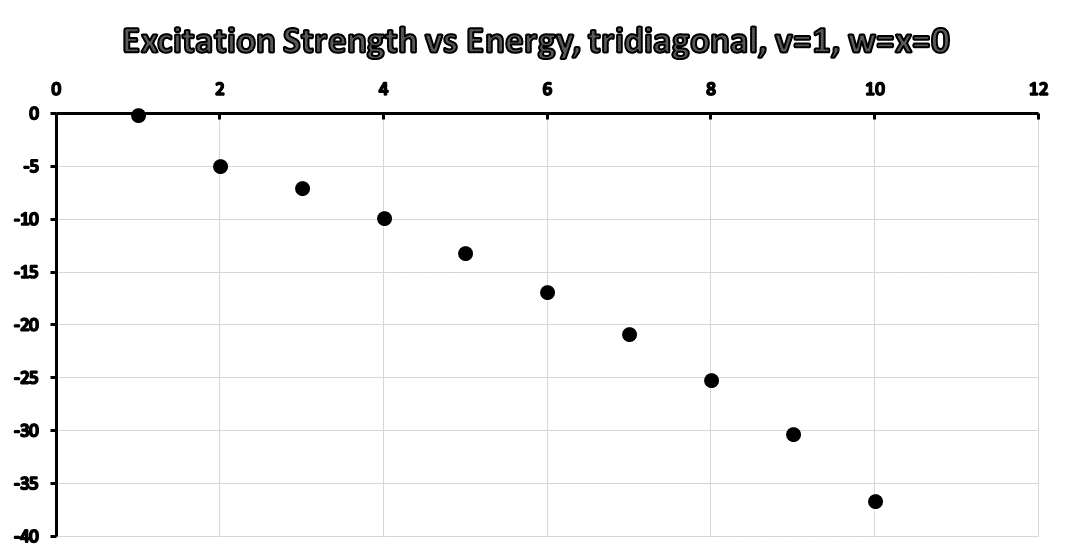}}{}
\caption*{}
\end{figure}

\end{center}

\begin{center}
    
\begin{figure}[!htbp]%
\footnotesize
\stackunder[5pt]{\includegraphics[width=6cm,height=3.66cm]{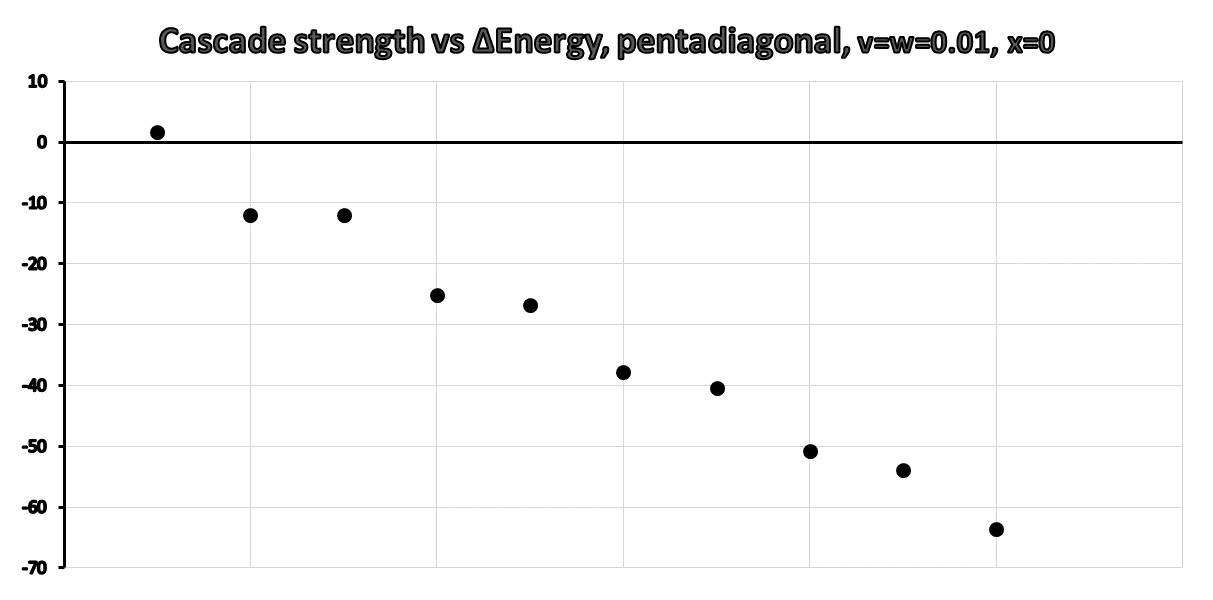}}{ }%
\hspace{1cm}%
\stackunder[5pt]{\includegraphics[width=6cm,height=3.66cm]{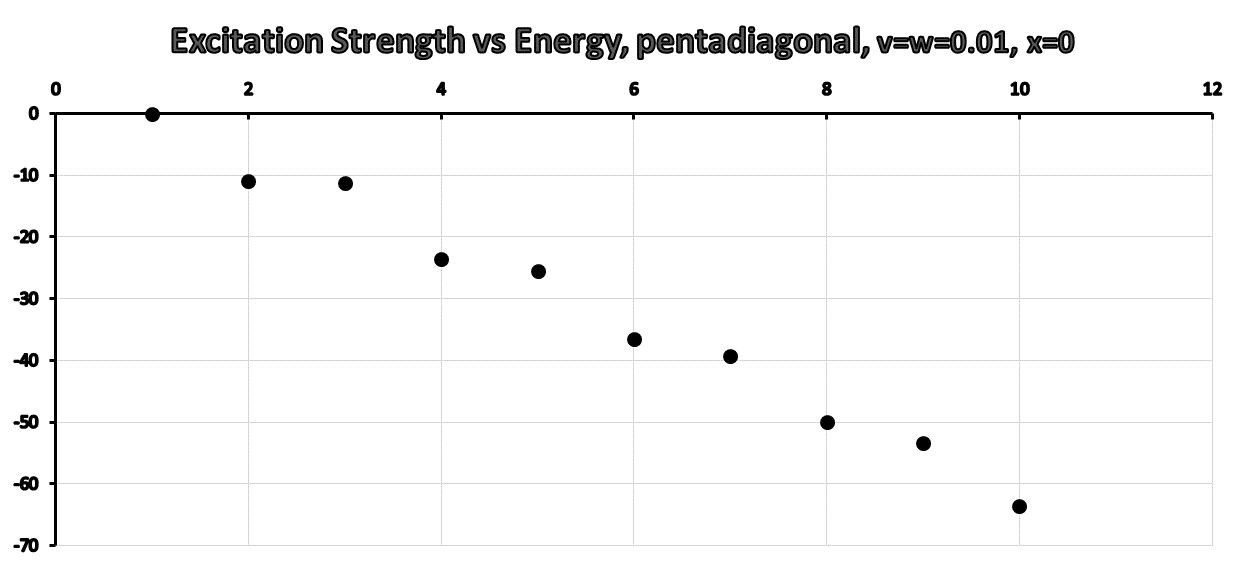}}{}
\caption*{}
\end{figure}

\end{center}

\begin{center}
    
\begin{figure}[!htbp]%
\footnotesize
\stackunder[5pt]{\includegraphics[width=6cm,height=3.66cm]{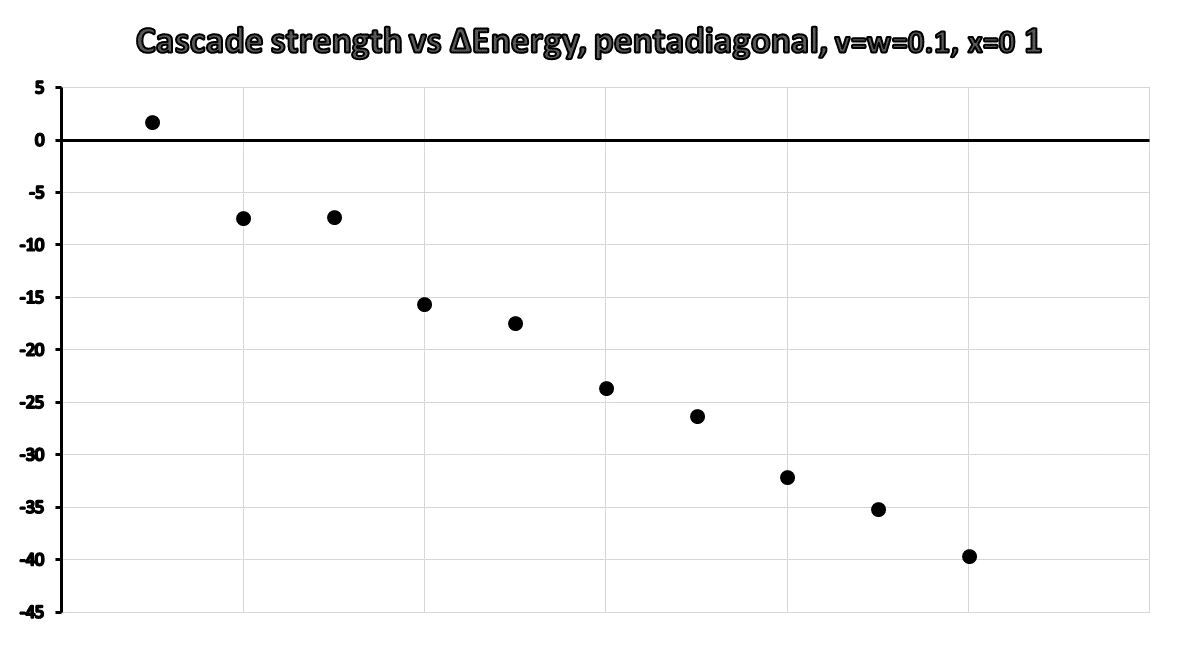}}{ }%
\hspace{1cm}%
\stackunder[5pt]{\includegraphics[width=6cm,height=3.66cm]{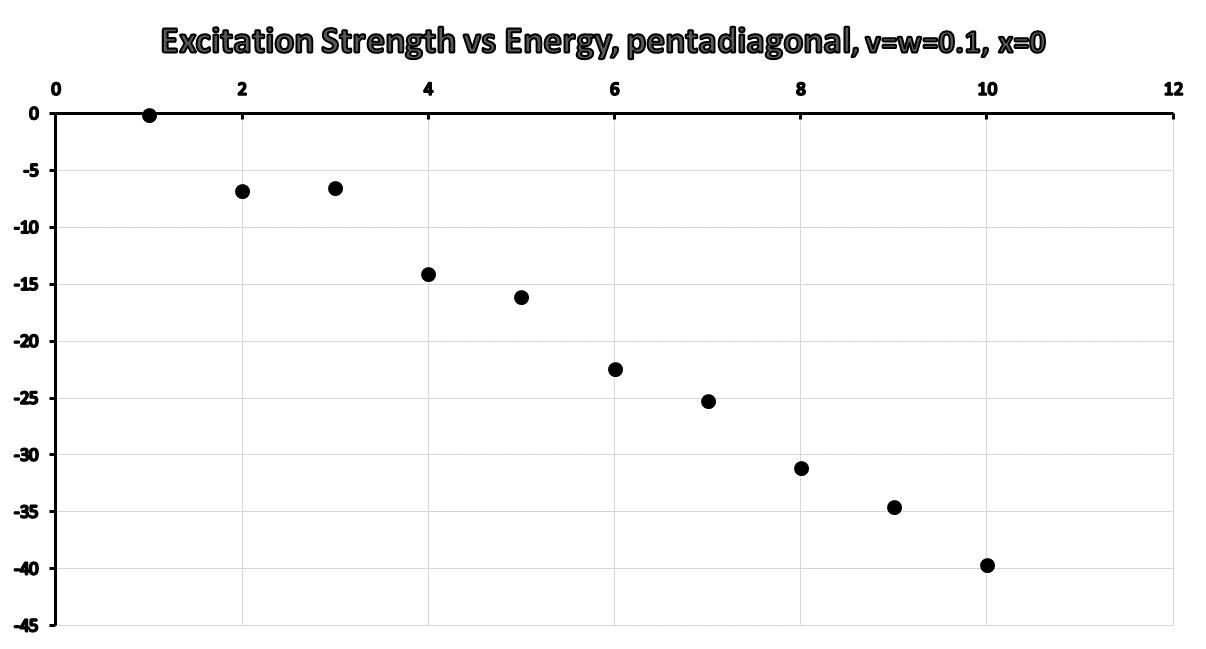}}{}
\caption*{}
\end{figure}

\end{center}

\begin{center}
    
\begin{figure}[!htbp]%
\footnotesize
\stackunder[5pt]{\includegraphics[width=6cm,height=3.66cm]{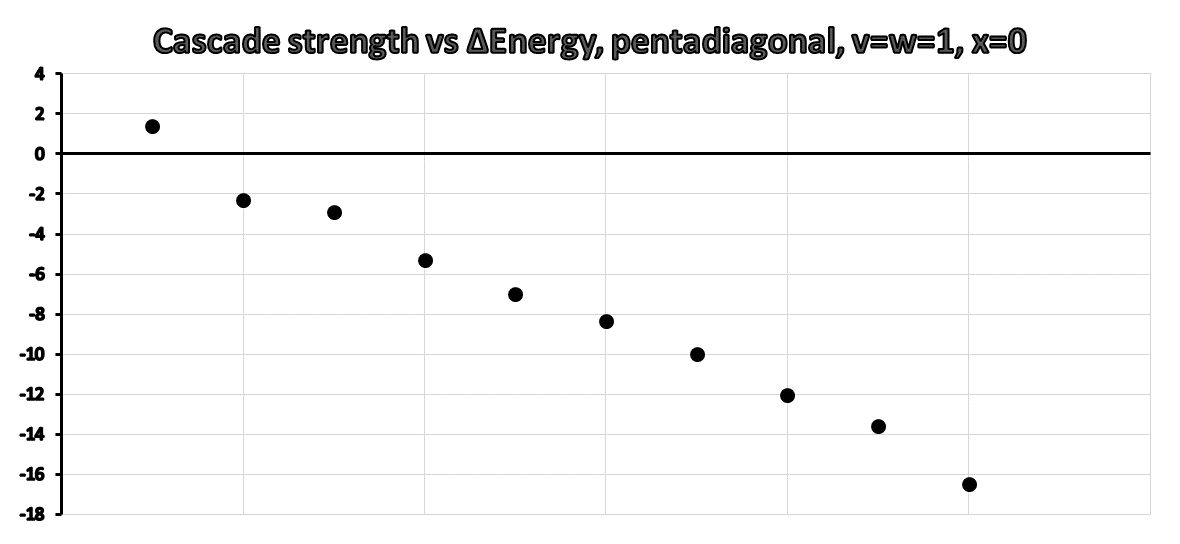}}{ }%
\hspace{1cm}%
\stackunder[5pt]{\includegraphics[width=6cm,height=3.66cm]{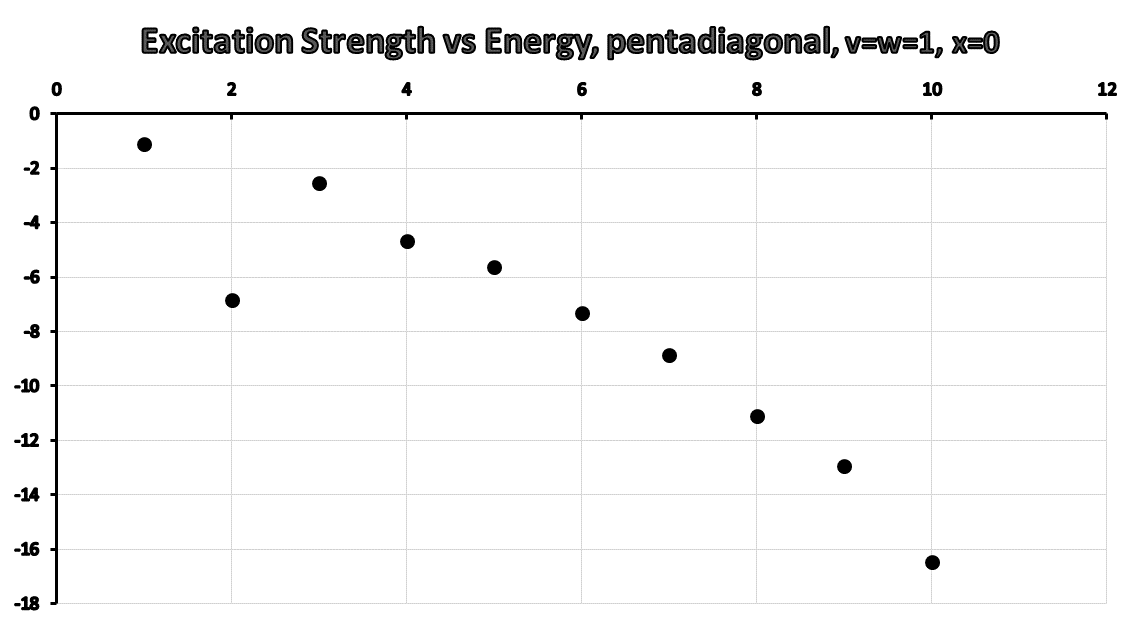}}{}
\caption*{}
\end{figure}

\end{center}

\begin{center}
    
\begin{figure}[!htbp]%
\footnotesize
\stackunder[5pt]{\includegraphics[width=6cm,height=3.66cm]{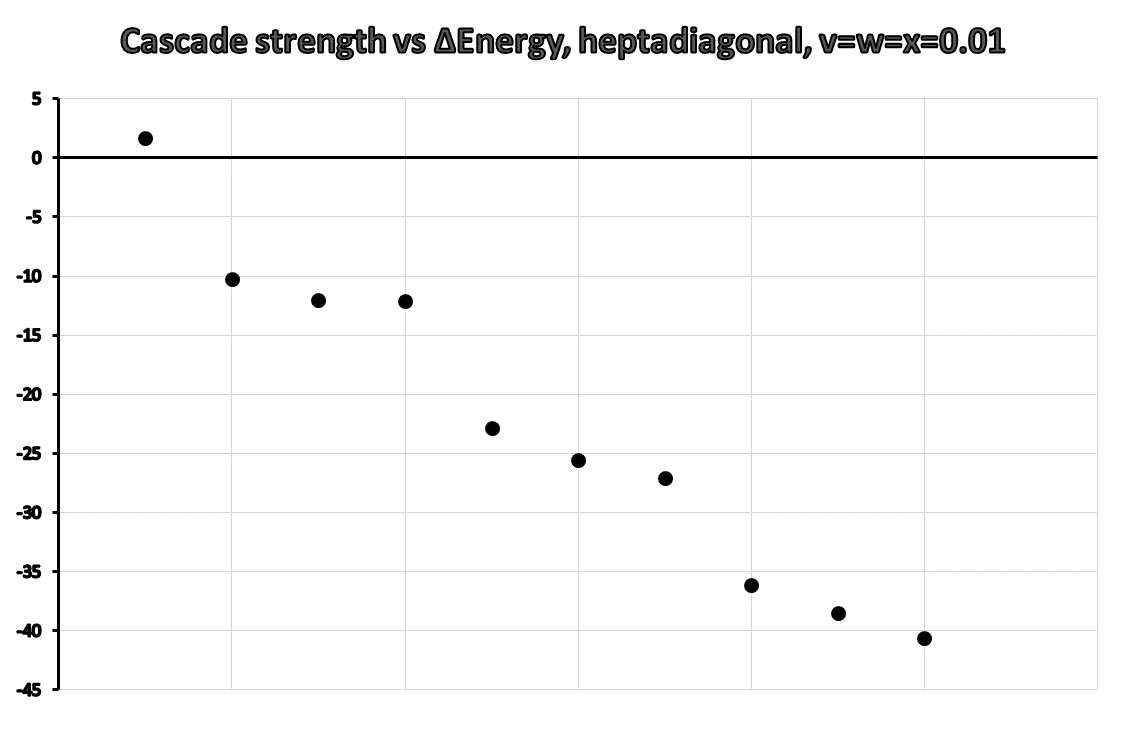}}{ }%
\hspace{1cm}%
\stackunder[5pt]{\includegraphics[width=6cm,height=3.66cm]{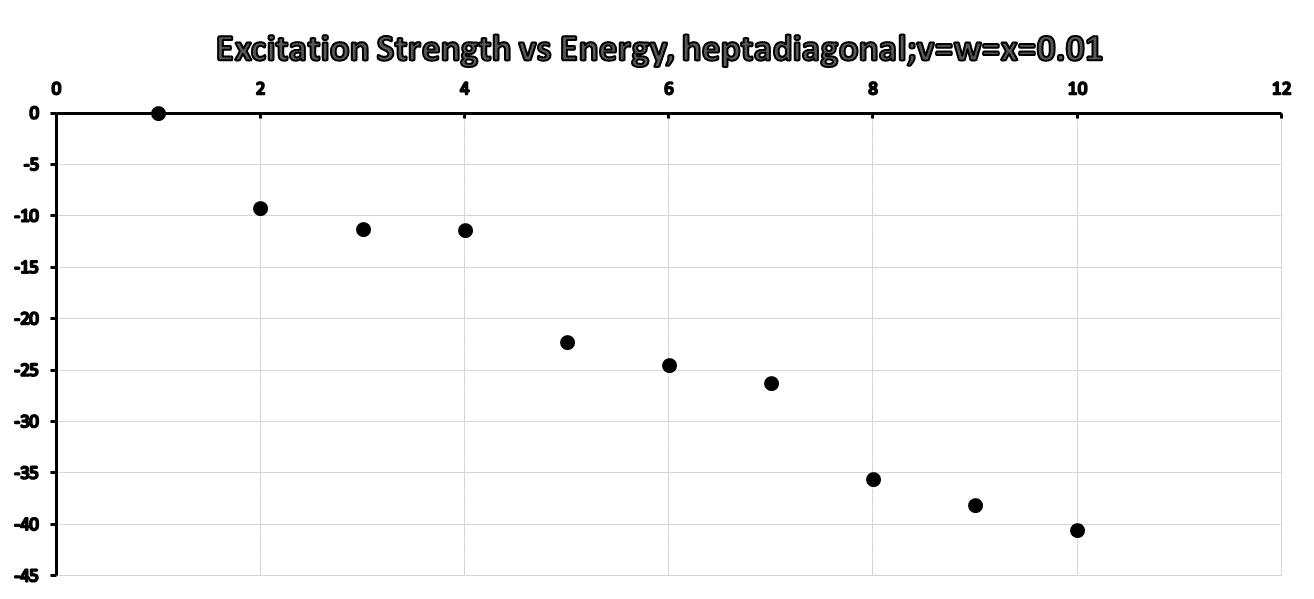}}{}
\caption*{}
\end{figure}

\end{center}

\begin{center}
    
\begin{figure}[!htbp]%
\footnotesize
\stackunder[5pt]{\includegraphics[width=6cm,height=3.66cm]{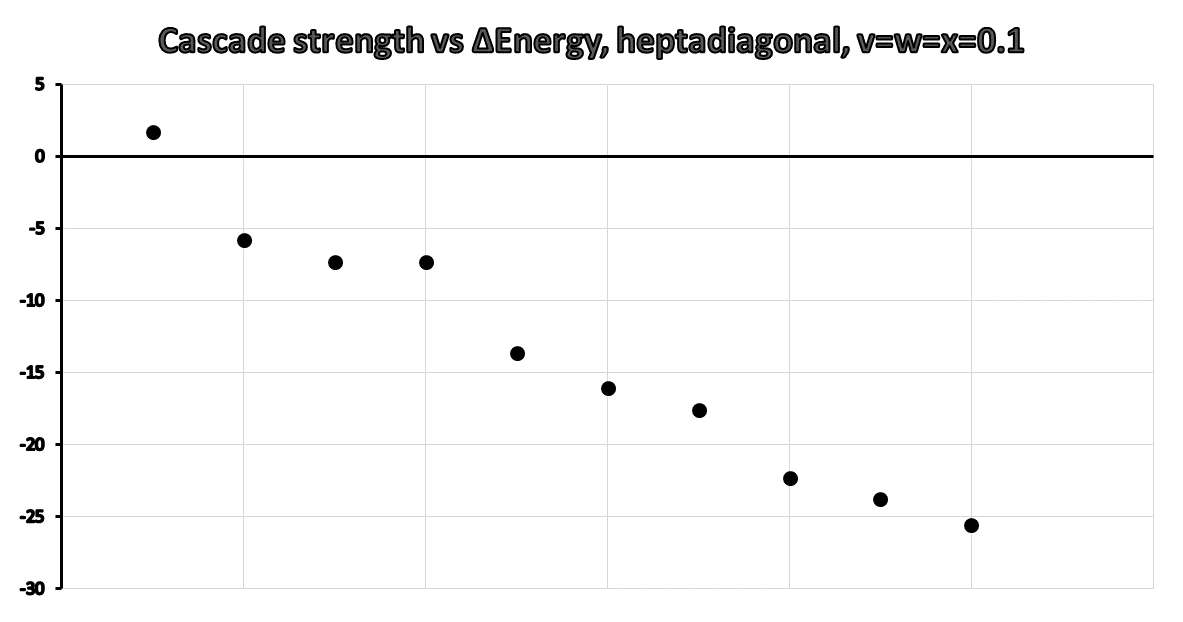}}{ }%
\hspace{1cm}%
\stackunder[5pt]{\includegraphics[width=6cm,height=3.66cm]{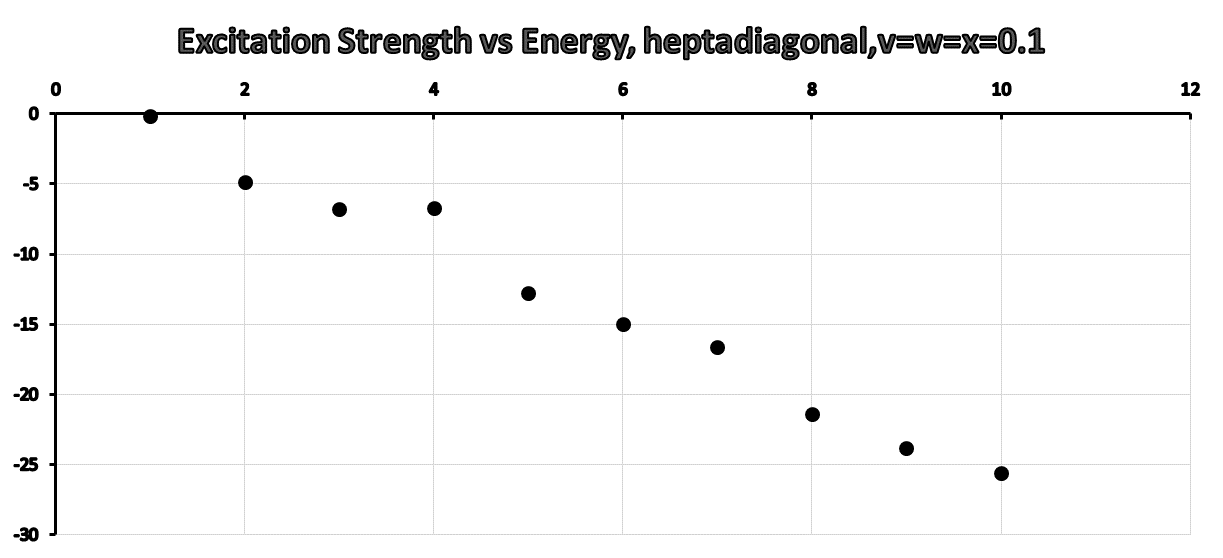}}{}
\caption*{}
\end{figure}

\end{center}

\begin{center}
    
\begin{figure}[!htbp]%
\footnotesize
\stackunder[5pt]{\includegraphics[width=6cm,height=3.66cm]{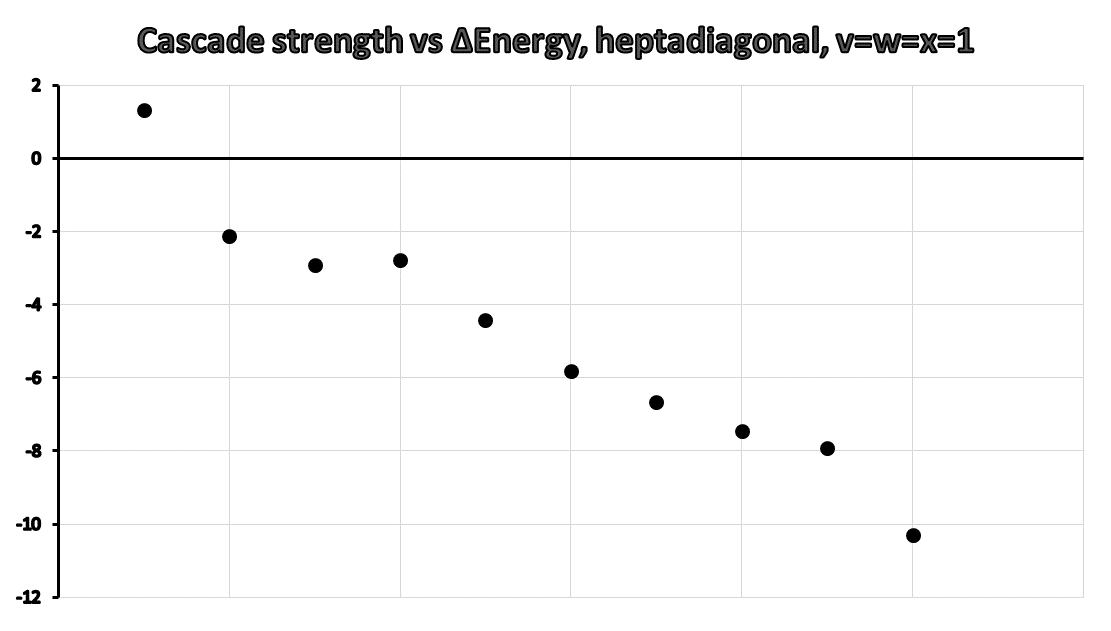}}{ }%
\hspace{1cm}%
\stackunder[5pt]{\includegraphics[width=6cm,height=3.66cm]{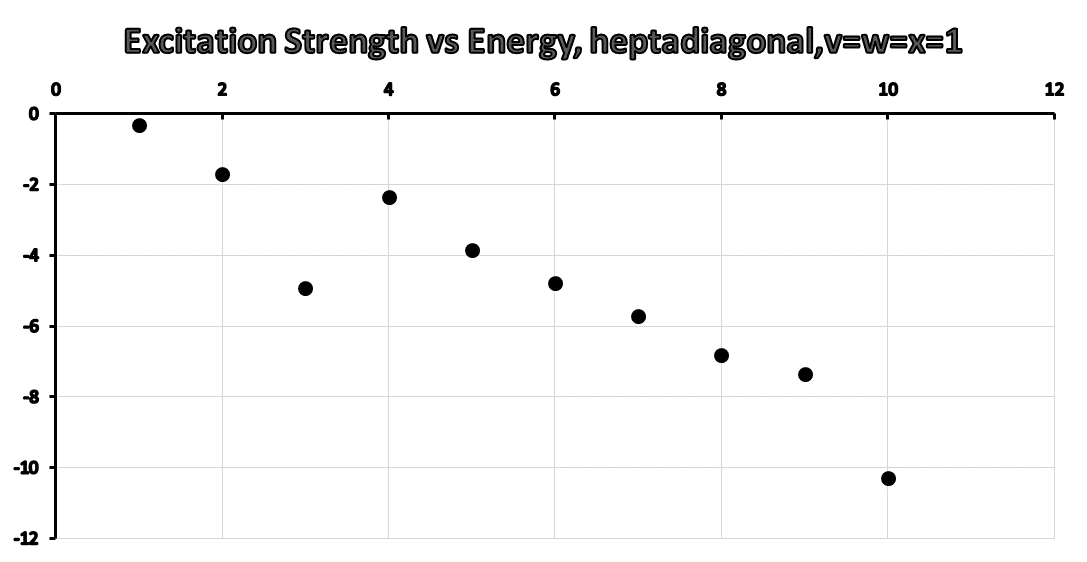}}{}
\caption*{}
\end{figure}

\end{center}

\begin{center}
    
\begin{figure}[!htbp]%
\footnotesize
\stackunder[5pt]{\includegraphics[width=6cm,height=3.66cm]{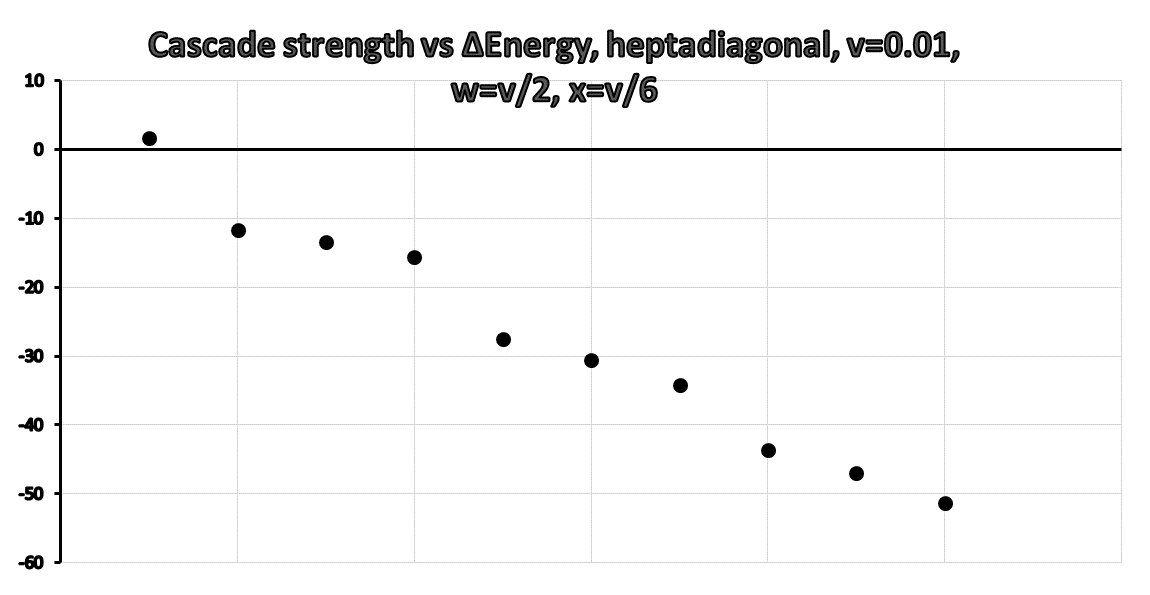}}{ }%
\hspace{1cm}%
\stackunder[5pt]{\includegraphics[width=6cm,height=3.66cm]{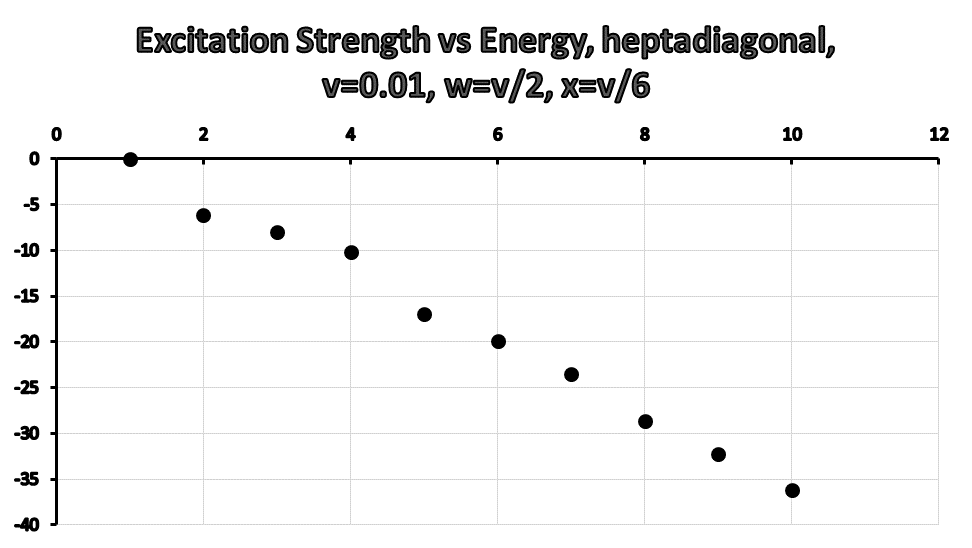}}{}
\caption*{}
\end{figure}

\end{center}

\begin{center}
    
\begin{figure}[!htbp]%
\footnotesize
\stackunder[5pt]{\includegraphics[width=6cm,height=3.66cm]{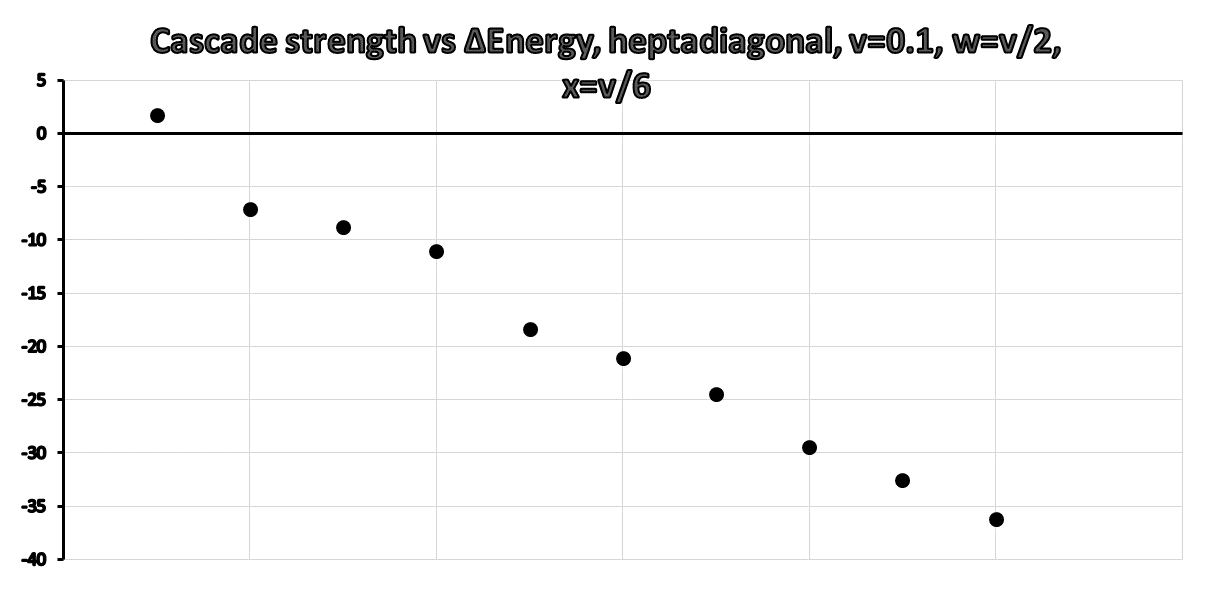}}{ }%
\hspace{1cm}%
\stackunder[5pt]{\includegraphics[width=6cm,height=3.66cm]{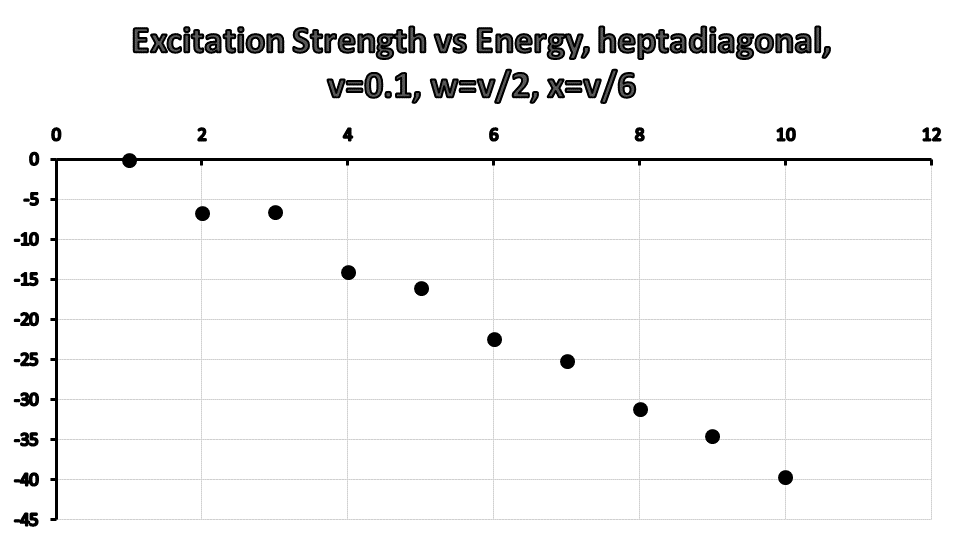}}{}
\caption*{}
\end{figure}

\end{center}

\begin{center}
    
\begin{figure}[!htbp]%
\footnotesize
\stackunder[5pt]{\includegraphics[width=6cm,height=3.66cm]{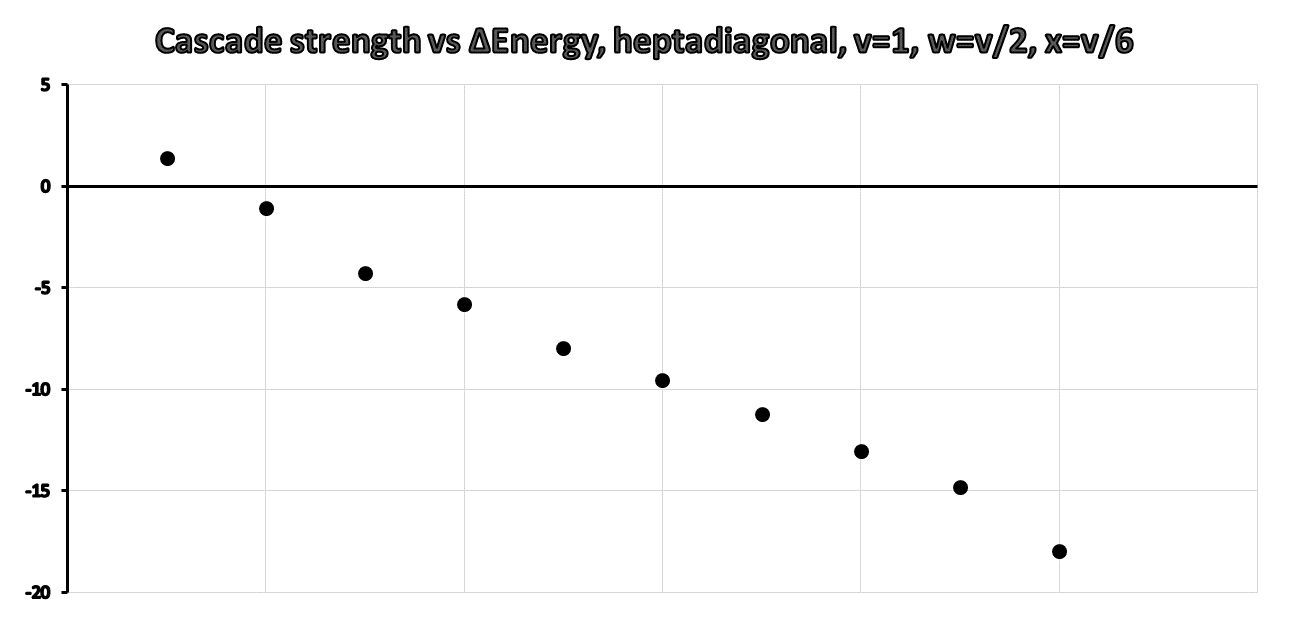}}{ }%
\hspace{1cm}%
\stackunder[5pt]{\includegraphics[width=6cm,height=3.66cm]{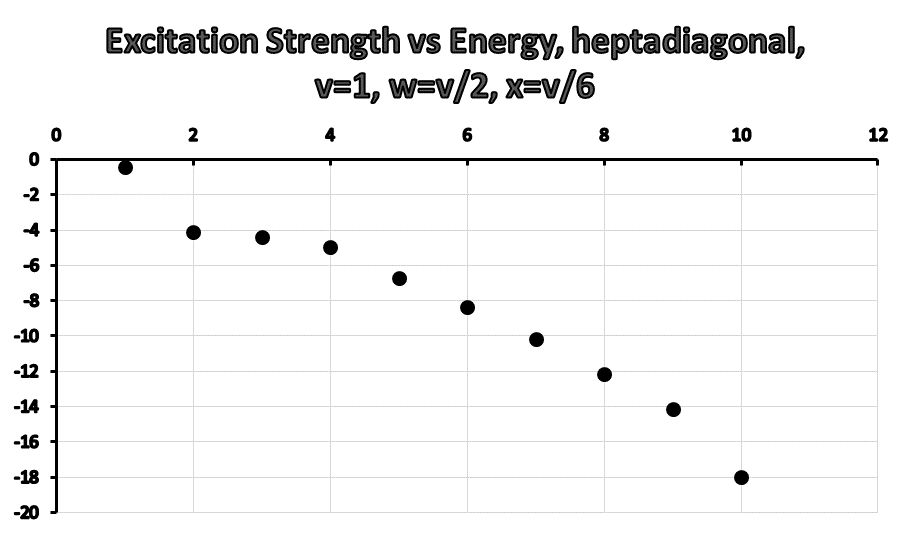}}{}
\caption*{}
\end{figure}

\end{center}

\section{Discussion}

We have put the cascade and excition figures side by side to make
the point that , although not identical, they are remarkably similar.
We emphasize that cascade is not the reverse of excitation. In the latter we always start
with the lowest state and go to the n'th excited state. In cascade we have many other
(reverse) transitions e.g. from the eighth excited state to the fifth one.
We first consider the tridiagonal case. One gets the best linear behavior for
v/E=0.1.Linearity on a log plot implies exponential behavior. For v/E=1 the curve is
dominantly linear but there is a slight arch. Surprisingly for v=0.01, there is a kink at n=9.
We will later see that this is not the case for asymptotically small v/E.
In the pentadiagonal case for transition strength for v=1 there is a large dip at n=2-
destructive interference. It is present but less pronounced in the cascade case; also
present but less pronounced for smaller v/E.
For the heptadiagonal transition case with v=w=x=1 we also get a pronounced dip
but at n=3. In the cascade case there is linear behavior from n=4 onward but the curve
flattens out for n= 4, 3, and 2 but here is then a significant rise for n=1. However when,
again for the heptadiagonal case we set v=1, w=$\frac{1}{2!}$, x=$\frac{1}{3!}$ the cascade curve looks quite linear again. This is not a complete surprise because when we make w and x smaller, we go in the direction of the tridiagonal result.
In Tables 2, 3 and 4 we show the excitation strengths for v=0.01, 0.1 and in tables
5, 6, and 7 the corresponding cascade averages. In contrast to the figures we here
show side by side the tri,penta and hepta diagonal results for a given v.

\begin{center}
\begin{table}[H]
\captionof{table}{Excitation Spectrum for v=1} %
\begin{tabular}{|l|l|l|l|l|}
\hline 
 & tri  & penta  & hepta  & hepta vwx \tabularnewline
\hline 
1  & -0.17879  & -1.09417  & -0.3195  & -0.41621 \tabularnewline
\hline 
2  & -4.93803  & -6.83181  & -1.71336  & -4.11011 \tabularnewline
\hline 
3  & -7.07651  & -2.52695  & -4.92044  & -4.37961 \tabularnewline
\hline 
4  & -9.88843  & -4.66744  & -2.33968  & -4.94979 \tabularnewline
\hline 
5  & -13.1961  & -5.62655  & -3.85475  & -6.69694 \tabularnewline
\hline 
6  & -16.8808  & -7.32741  & -4.78282  & -8.36112 \tabularnewline
\hline 
7  & -20.8785  & -8.8552  & -5.72363  & -10.1684 \tabularnewline
\hline 
8  & -25.2209  & -11.1213  & -6.80623  & -12.1534 \tabularnewline
\hline 
9  & -30.338  & -12.9337  & -7.35953  & -14.1213 \tabularnewline
\hline 
10  & -36.6644  & -16.4684  & -10.2962  & -17.9763 \tabularnewline
\hline 
\end{tabular}
\end{table}

\par\end{center}

\begin{center}
\begin{table}[H]
\captionof{table}{Excitation Spectrum for v=0.1} %
\begin{tabular}{|l|l|l|l|l|}
\hline 
 & tri  & penta  & hepta  & hepta vwx \tabularnewline
\hline 
1  & -0.00738  & -0.14681  & -0.12787  & -0.14681 \tabularnewline
\hline 
2  & -6.41019  & -6.7705  & -4.84763  & -6.7705 \tabularnewline
\hline 
3  & -15.2755  & -6.55817  & -6.75191  & -6.55817 \tabularnewline
\hline 
4  & -23.5227  & -14.13  & -6.68711  & -14.13 \tabularnewline
\hline 
5  & -31.8735  & -16.099  & -12.7848  & -16.099 \tabularnewline
\hline 
6  & -40.4297  & -22.4111  & -14.9645  & -22.4111 \tabularnewline
\hline 
7  & -49.2041  & -25.2178  & -16.6326  & -25.2178 \tabularnewline
\hline 
8  & -58.1883  & -31.1712  & -21.4201  & -31.1712 \tabularnewline
\hline 
9  & -67.3691  & -34.5585  & -23.8361  & -34.5585 \tabularnewline
\hline 
10  & -76.7596  & -39.6674  & -25.5686  & -39.6674 \tabularnewline
\hline 
\end{tabular}
\end{table}

\par\end{center}

\begin{center}
\begin{table}[H]
\captionof{table}{Excitation Spectrum for v=0.01} %
\begin{tabular}{|l|l|l|l|l|}
\hline 
 & tri  & penta  & hepta  & hepta vwx \tabularnewline
\hline 
1  & -7.6E-05  & -0.01423  & -0.01404  & -0.07396 \tabularnewline
\hline 
2  & -10.9735  & -11.009  & -9.24842  & -6.13671 \tabularnewline
\hline 
3  & -24.486  & -11.2124  & -11.2329  & -8.00813 \tabularnewline
\hline 
4  & -37.3515  & -23.6196  & -11.3943  & -10.221 \tabularnewline
\hline 
5  & -50.3143  & -25.5145  & -22.2603  & -16.9359 \tabularnewline
\hline 
6  & -63.4797  & -36.5333  & -24.4962  & -19.9302 \tabularnewline
\hline 
7  & -76.8619  & -39.3634  & -26.2338  & -23.4957 \tabularnewline
\hline 
8  & -90.4727  & -49.9249  & -35.549  & -28.6644 \tabularnewline
\hline 
9  & -94.8711  & -53.4074  & -38.0957  & -32.2198 \tabularnewline
\hline 
10  & -118.209  & -63.6203  & -40.5783  & -36.1814 \tabularnewline
\hline 
\end{tabular}
\end{table}

\par\end{center}

In Table 5-7 we show results for the cascade decays.

\begin{table}[H]
\captionof{table}{Cascade Decays for v=1} %
\begin{tabular}{|l|l|l|l|l|}
\hline 
 & tri  & penta  & hepta  & hepta vwx \tabularnewline
\hline 
1  & 1.475208  & 1.375723  & 1.33565  & 1.419713 \tabularnewline
\hline 
2  & -2.97858  & -2.28142  & -2.12546  & -1.07107 \tabularnewline
\hline 
3  & -7.35646  & -2.92052  & -2.91908  & -4.28074 \tabularnewline
\hline 
4  & -10.701  & -5.31517  & -2.782  & -5.81003 \tabularnewline
\hline 
5  & -14.253  & -6.98987  & -4.42519  & -7.94482 \tabularnewline
\hline 
6  & -17.9651  & -8.33056  & -5.80058  & -9.54138 \tabularnewline
\hline 
7  & -21.9022  & -9.99003  & -6.65594  & -11.2324 \tabularnewline
\hline 
8  & -26.0397  & -12.055  & -7.46434  & -13.0221 \tabularnewline
\hline 
9  & -30.8862  & -13.58  & -7.90574  & -14.7925 \tabularnewline
\hline 
10  & -36.6644  & -16.4684  & -10.2962  & -17.9763 \tabularnewline
\hline 
\end{tabular}
\end{table}

\begin{table}[H]
\captionof{table}{Cascade Decays for v=0.1} %
\begin{tabular}{|l|l|l|l|l|}
\hline 
 & tri  & penta  & hepta  & hepta vwx \tabularnewline
\hline 
1  & 1.700507  & 1.699669  & 1.698715  & 1.700062 \tabularnewline
\hline 
2  & -7.36563  & -7.41067  & -5.80332  & -7.1114 \tabularnewline
\hline 
3  & -16.9479  & -7.37501  & -7.3255  & -8.75926 \tabularnewline
\hline 
4  & -25.274  & -15.6679  & -7.37551  & -11.0702 \tabularnewline
\hline 
5  & -33.555  & -17.4211  & -13.6674  & -18.3371 \tabularnewline
\hline 
6  & -41.9662  & -23.6271  & -16.0857  & -21.1241 \tabularnewline
\hline 
7  & -50.5406  & -26.2587  & -17.5985  & -24.4477 \tabularnewline
\hline 
8  & -59.2485  & -32.1169  & -22.3293  & -29.4771 \tabularnewline
\hline 
9  & -68.0405  & -35.2084  & -23.7897  & -32.5574 \tabularnewline
\hline 
10  & -76.7596  & -39.6674  & -25.5686  & -36.1814 \tabularnewline
\hline 
\end{tabular}
\end{table}

\begin{table}[H]
\captionof{table}{Cascade Decays for v=0.01} %
\begin{tabular}{|l|l|l|l|l|}
\hline 
 & tri  & penta  & hepta  & hepta vwx \tabularnewline
\hline 
1  & 1.704704  & 1.704701  & 1.704698  & 1.704703 \tabularnewline
\hline 
2  & -11.9639  & -11.969  & -10.2798  & -11.738 \tabularnewline
\hline 
3  & -26.1929  & -12.0361  & -12.0326  & -13.491 \tabularnewline
\hline 
4  & -39.1037  & -25.0972  & -12.081  & -15.6637 \tabularnewline
\hline 
5  & -51.9939  & -26.7729  & -22.833  & -27.57 \tabularnewline
\hline 
6  & -65.0163  & -37.8633  & -25.5572  & -30.5178 \tabularnewline
\hline 
7  & -78.1967  & -40.4355  & -27.0735  & -34.2006 \tabularnewline
\hline 
8  & -79.6778  & -50.8625  & -36.0801  & -43.6991 \tabularnewline
\hline 
9  & -89.5096  & -53.9601  & -38.4745  & -47.0518 \tabularnewline
\hline 
10  & -118.209  & -63.6203  & -40.5783  & -51.3127 \tabularnewline
\hline 
\end{tabular}
\end{table}

\section{Weak Coupling Limit for a Tridiagonal Matrix with T$_{2}$ Transition
Operator}

We can show that with T$_2$ tranistion operator \textless{}n T$_{2}$
(n+1)\textgreater{} = $\sqrt{n+1}$

in the weak coupling limit $u=\frac{v}{E}$ is very small

O(0 $\rightarrow$ n) has the structure $(\frac{v}{E})^{(n-1)}$ A$_{n}$
where A$_{n}$ does not depend on v/E.

ln (O$^{2}$)= (n-1) ln($(\frac{v}{E})^{2}$) +ln (A$_{n}^{2}$) in
the weak coupling limit

Note that by making $\frac{v}{E}$ very small ln(($\frac{v}{E}$) $^{2}$)
becomes very large and negative thus drowning out the term ln (A$_{n}^{2}$).
Thus we get ln($O^{2}$) approaches (n-1) ln($(\frac{v}{E})^{2}$).
The fact that this ln($O^{2}$) is linear in (n-1) proves that we
have exponential behavior for $O^{2}$. In a previous work {[}5?{]}
we considered the weak coupling limit for the transition operator
T$_1$. In that case O(1 $\rightarrow$ n) was of the $(\frac{v}{E})^{m}$
A(1 $\rightarrow$ n). However m was not the same as (n-1) so there
was no exponential behavior for T$_1$. As shown in ref {[}5{]} the values
of m from 1 to 9 are respectively 0, 3, 4, 5, 6, 5, 6, 7, 8. For n=10
the value of O is zero (ln($O^2$) = $- \infty$).

\captionof{table}{T2 Transition Strength, $\frac{v}{E}=0.0001$} %
\begin{tabular}{|c|c|c|c|c|}
\hline 
0$\rightarrow n$  & ln (A$_{n}^{2}$)  & (n-1) ln ($(\frac{v}{E})^{2}$)  & sum &Exact\tabularnewline
\hline 
1  & 0  & 0  & 0&0\tabularnewline
\hline 
2  & -1.7627  & -18.42  & -20.1034&-20.18\tabularnewline
\hline 
3  & -6.0653  & -36.84  & -42.9060&-42.91\tabularnewline
\hline 
4  & -9.7206  & -55.26  & -64.9827&-64.92\tabularnewline
\hline 
5  & -13.4731  & -73.68  & -87.1558&-87.16\tabularnewline
\hline 
6  & -9.5394  & -92.103  & -101.6340&-101.63\tabularnewline
\hline 
7  & -9.1366  & -110.52  & -119.6610&-119.36\tabularnewline
\hline 
8  & -10.4230  & --128.945  & -139.368&-139.36\tabularnewline
\hline 
9  & -12.6063  & -147.37  & -159.972&-159.97\tabularnewline
\hline 
10  & -35.3138  & -165.79  & -201.102&-201.1\tabularnewline
\hline 
\end{tabular}%

For the T$_2$ transition operator the expressions for A$_n$ are given in
Table 10 in terms of g$_k$=$\frac{1}{k!}$

\pagebreak

\captionof{table}{T1 Transition Strength, $\frac{v}{E}=0.0001$}
\begin{tabular}{|c|c|c|c|c|c|}
\hline 
n & m & ln(A$_{n}^{2}$) & ln($\frac{v}{E}$)$^{m}$ & Sum & Exact\tabularnewline
\hline 
1 & 0 & 0 & 0 & 0 & 0\tabularnewline
\hline 
2 & 3 & -0.8109 & -55.2620 & -56.07 & -57.46\tabularnewline
\hline 
3 & 4 & -4.1589 & -73.6827 & -77.84 & -80.04\tabularnewline
\hline 
4 & 5 & -5.9915 & -92.1034 & -98.09 & -102.48\tabularnewline
\hline 
5 & 6 & -8.3627 & -110.5240 & -118.89 & -116.49\tabularnewline
\hline 
6 & 5 & -9.5750 & -92.1034 & -101.67 & -101.62\tabularnewline
\hline 
7 & 6 & -9.9396 & -110.5241 & -120.46 & -120.46\tabularnewline
\hline 
8 & 7 & -11.6342 & -128.944 & -140.46 & -140.58\tabularnewline
\hline 
9 & 8 & -14.0985 & -147.3654 & -161.46 & -161.46\tabularnewline
\hline 
10 &  &  &  &  & $-\infty$ \tabularnewline
\hline 
\end{tabular}

\captionof{table}{ Expressions for A$_{n}$ for T$_2$ transition
operator g$_{k}$=$\frac{1}{k!}$ .} 
\begin{table}[H]
\begin{tabular}{|l|l|}
\hline 
A$_{n}$ & expression in terms of g$_k$ \tabularnewline
\hline 
1  & 1 \tabularnewline
\hline 
2  & 1-$\sqrt{2}$\tabularnewline
\hline 
3  & g$_2$-$\sqrt{2}$+g$_2${*}$\sqrt{3}$ \tabularnewline
\hline 
4  & g$_3$-g$_2${*}$\sqrt{2}$+g$_2${*}$\sqrt{3}$-g$_3${*}$\sqrt{4}$ \tabularnewline
\hline 
5  & g$_4$-g$_3${*}$\sqrt{2}$+g$_2${*}g$_2${*}$\sqrt{3}$-g$_3${*}$\sqrt{4}$+g$_4${*}$\sqrt{5}$\tabularnewline
\hline 
6  & -g$_4${*}$\sqrt{2}$+g$_2${*}g$_3${*}($\sqrt{3}$-$\sqrt{4}$)+g$_4${*}$\sqrt{5}$-g$_5${*}$\sqrt{6}$\tabularnewline
\hline 
7  & g$_2${*}g$_4${*}$\sqrt{3}$-g$_3${*}g$_3${*}$\sqrt{4}$+g$_4${*}g$_2${*}$\sqrt{5}$-g$_5${*}$\sqrt{6}$+g$_6${*}$\sqrt{7}$\tabularnewline
\hline 
8  & g$_4${*}g$_3${*}($\sqrt{5}$-$\sqrt{4}$)-g$_2${*}g$_5${*}$\sqrt{6}$ +g$_6${*}$\sqrt{7}$-g$_7${*}$\sqrt{8}$\tabularnewline
\hline 
9  & %
\begin{tabular}{@{}l@{}}
g$_4${*}g$_4${*}$\sqrt{5}$-g$_5${*}g$_3${*}$\sqrt{6}$+g$_6${*}g$_2${*}$\sqrt{7}$-g$_7${*}$\sqrt{8}$+g$_8${*}$\sqrt{9}$\tabularnewline
\end{tabular}\tabularnewline
\hline 
10 & g$_9$-g$_8${*}$\sqrt{2}$+g$_2${*}g$_7${*}$\sqrt{3}$-g$_6${*}g$_3${*}$\sqrt{4}$+g$_4${*}g$_5${*}$\sqrt{5}$-g$_4${*}g$_5${*}$\sqrt{6}$+g$_6${*}g$_3${*}$\sqrt{7}$-g$_7${*}g$_2${*}$\sqrt{8}$+g$_8${*}$\sqrt{9}$-g$_9${*}$\sqrt{10}$ \tabularnewline
\hline 
\end{tabular}
\end{table}

In the (n,$\gamma$) results the best example of a linear behavior on a log plot is given
in Fig 1 of Brown and Larsen [7]. Of course the Hamiltonian they use is completely
different from the simple one used here, but it suggests that the phenomenon of
exponential behavior is more widespread than one might initially think. We have given
support to this notion by proving that one gets exponential behavior in the weak
coupling limit.


\begin{thebibliography}{10}
\bibitem{key-1} A.Kingan, M. Quinonez, X. Yu and L. Zamick , International
Journal of Modern Physics E Vol. 28,1850090 (2018)

\bibitem{key-2}A.Kingan, M. Quinonez, X. Yu and L. Zamick , arXiv:nucl-th/1803.00645
(2018)

\bibitem{key-1}A. Kingan and L. Zamick, International Journal of
Modern Physics E, Vol 26,(2018) 1850064

\bibitem{key-5}A.Kingan and L. Zamick ,International Journal of Modern
Physics E Vol.27, NO 10 (2018) 1850087

\bibitem{key-6}L.Wolfe and L. Zamick, International Journal of Modern
Physics E Vol. 28, No. 5 (2019) 1950037

\bibitem{key-4}R. Schwengner, S. Frauendorf and A. C. Larson, Phys.
Rev. Lett. 111 (2013) 232504

\bibitem{key-5}B. Alex Brown and A.C. Larsen Phys. Rev. Lett. 113,
252502 (2014)

\bibitem{key-9}Siega, Phys. Rev. Lett. 119 (2017) 052502 Karampagia,
B. A. Brown and V. Zelevinsky, Phys. Rev C 95 (2017) 024322

\bibitem{key-11}R. Schwengner, S. Frauendorf and B.A. Brown, Phys.
Rev. Lett. 118 (2017) 092507

\bibitem{key-10}R. Schwengner, S. Frauendorf and A. C. Larson, Phys.
Rev. Lett. 111 (2013) 232504 B. Alex Brown and A.C. Larsen Phys. Rev.
Lett. 113, 252502 (2014)

\bibitem{key-12}Siega, Phys. Rev. Lett. 119 (2017) 052502 Karampagia,
B. A. Brown and V. Zelevinsky, Phys. Rev C 95 (2017) 024322

\bibitem{key-14}R. Schwengner, S. Frauendorf and B.A. Brown, Phys.
Rev. Lett. 118 (2017) 092507\end{thebibliography}
\end{document}